# Spin-orbit torque induced dipole skyrmion motion at room temperature

Sergio A. Montoya[1], Robert Tolley[2,3], Ian Gilbert[4], Soong-Geun Je[5,6], Mi-Young Im[5,6], Eric E. Fullerton[2,3]

[1]*Space and Naval Warfare Systems Center Pacific, San Diego, CA 92152, USA*
[2]*Center for Memory and Recording Research, University of California, San Diego, La Jolla, CA 92093, USA*
[3]*Department of Electrical and Computer Engineering, University of California, La Jolla, CA 92093, USA*
[4]*Center for Nanoscale Science and Technology, National Institute of Standards and Technology, Gaithersburg, Maryland 20899, USA*
[5]*Center for X-ray Optics, Lawrence Berkeley National Laboratory, Berkeley, California 94720, USA*
[6]*Department of Emerging Materials Science, Daegu Gyeongbuk Institute of Science and Technology, Daegu, Korea*

Dated: May 30, 2018

**Abstract**

We demonstrate deterministic control of dipole-field-stabilized skyrmions by means of spin-orbit torques arising from heavy transition-metal seed layers. Experiments are performed on amorphous Fe/Gd multilayers that are patterned into wires and exhibit stripe domains and dipole skyrmions at room temperature. We show that while the domain walls and skyrmions are achiral on average due to lack of Dzyaloshinskii-Moriya interactions, the Néel-like closure domain walls at each surface are chiral and can couple to spin-orbit torques. The current-induced domain evolutions are reported for different magnetic phases, including disordered stripe domains, coexisting stripes and dipole skyrmions and a closed packed dipole skyrmion lattice. The magnetic textures exhibit motion under current excitations with a current density $\sim 10^8$ A/m$^2$. By comparing the motion resulting from magnetic spin textures in Fe/Gd films with different heavy transition-metal interfaces, we confirm spin currents can be used to manipulate achiral dipole skyrmions via spin-orbit torques.

**Introduction.**

The ability to locally control magnetism through the use of electrical currents has led to the development of a range of new non-volatile and efficient magnetism-based information technologies [1-3]. Current-induced magnetization dynamics were initially observed from spin-transfer torques in nanopillar devices with two magnetic layers [4-7]. The current becomes spin polarized by transmission through or upon reflection from the first magnetic layer and maintains this polarization as it passes through the non-magnetic spacer and interacts with the second ferromagnetic layer. This interaction provides a torque on the magnetic driving precessions that can result in steady-state magnetic dynamics or magnetization reversal [8-11]. A similar phenomenon occurs in magnetic wires where spin current exerts a torque on the domain wall that enables their displacement in the electron flow direction [12-14].

The discovery of spin-orbit torques (SOT) arising from for current-induced spin accumulation at the interface between a ferromagnetic layer and a non-magnetic layer with a large spin-orbit coupling (*e.g.* a heavy transition metal) provides an efficient method for manipulating spin structures within a single magnetic layer. The spin currents arise from the spin Hall effect in the non-magnetic metal layer when a charge current flows within the sample plane. The spin Hall effect generates a spin current that flows vertically towards the ferromagnet. When this spin current is absorbed in the ferromagnetic it applies a SOT. This torque is particularly effective at moving chiral domain walls or skyrmions stabilized by interfacial Dzyaloshinskii–Moriya interaction (DMI) that arise at the interface of the magnetic-nonmagnetic interface. The direction of motion is set by the helicity of the spin structure and the sign of the spin Hall angle in the non-magnetic layer. These observations demonstrated that magnetic spin textures could be displaced with low currents [15-22] in either patterned films [15-21] or continuous geometries [22]. These observations prompted research efforts to potentially develop next-generation high-efficiency and high-density information technologies in which individual skyrmions or chiral domain walls are used as logic and memory storage bits [23-27].

Thus far, the motion of chiral domain walls and skyrmions in thin films have been stabilized by interfacial DMI where the heavy-metal layer has the dual role of providing both the SOT and the DMI to stabilize the appropriate chiral spin structure. Here, we present direct observation of current-induced motion of magnetic stripe domains and dipole skyrmions in patterned wires of amorphous Fe/Gd multilayers at room temperature without the presence of DMI. The formation of sub-100nm dipole-stabilized biskyrmions and skyrmions in this class of material was previously reported in Refs. [28] and [29], respectively. These magnetic textures result solely from the competition between magneto-static and exchange energies [29] and there is no evidence of DMI in this system. Unlike skyrmions that form by means of DMI, a dipole skyrmion phase consists of an equal population of chiral cylindrical-like domains with two possible helicities (Fig. 1a, b) such that, on average, the material is achiral. This has been shown in Lorentz transmission electron microscopy which confirmed the dipole skyrmions possess a Bloch-like domain wall with equal population of two helicities for dipole-stabilized skyrmions in comparable Fe/Gd films [29].

In Fe/Gd multilayer films the dipole skyrmion phase becomes favorable for thick magnetic films when the thin-film shape anisotropy $2\pi M_S^2$ exceeds the uniaxial anisotropy $K_U$ so the ratio $Q = K_U / 2\pi M_S^2 < 1$ [30]. For such films, numerical simulations predicted that dipole skyrmions possess a domain wall that is Bloch-like at the center of the film but broadens and transitions to Néel-like walls towards the film surface (Figs. 1a, b) which are often referred to as closure domains or Néel caps. [31]. While the Bloch nature of wall in the center of the film will have one or two possible helicities shown in the Figs. 1a, b, the Néel-wall structure of the spins near the surfaces of the film have the same helicity for the top surface and the opposite helicity at the bottom surface (thus in total the spin structure is achiral). For the examples shown in Figs. 1a and b the magnetic spins at the top surface point inwards into the skyrmion whereas at the bottom surface they point outwards from the dipole skyrmion and this is independent of the helicity of the core. The chiral nature of the surface spins in closure domains, even in the absence of DMI, has previously been shown by x-ray resonant magnetic scattering from stripe domains using circularly polarized soft x-rays [32]. Here, we demonstrate that SOT can be used to manipulate dipole skyrmions, by interfacing the chiral surface spin structures of Fe/Gd films with heavy transition metal layers (*e.g.* Pt, Ta). Our results show the current-driven motion of dipole skyrmions depends, as expected, on the sign of the spin Hall angle of the transition metal layer and the chirality of the near-surface region of the stripe domains and dipole skyrmions.

**Results.**

**Domain morphology of Fe/Gd patterned wires.**

Figures 1c – 1h shows the magnetic domain structure of Fe/Gd multilayers films used in this study. Figures 1c – 1e are scanning electron microscopy with polarization analysis images of the stripe domains in remanence. This technique is surface sensitive and can image the projection of the magnetic spin x, y and z directions. Figures 1c-e detail the surface ($m_x$, $m_y$, $m_z$) magnetization distributions, where the perpendicular stripe domains with opposite orientation are represented along the ($m_z$) magnetization as dark/light contrast, whereas the Néel closure domains are solely seen by the periodic ($m_x$) magnetization consistent with the schematic in Figs. 1a and 1b. No contrast is observed along the ($m_y$) direction, as expected for stripe domains. This clearly shows the expected closure domain structure for the stripe phase. Because this imaging technique is sensitive to external magnetic fields we cannot image the surface domain structure in an applied field.

The field-dependent domain morphology and its current-induced motion were studied using full-field transmission soft x-ray microscopy measured at the Fe $L_3$ (708eV) absorption edge in 10-μm wide and 5-mm long Fe/Gd patterned wires (see Methods, for further details). Figures 1f-h shows the domain morphology of the Pt(5nm)/[Fe(0.34nm)/Gd(0.4nm)]x100/Ta(3nm) sample imaged while a perpendicular magnetic field was applied from negative to positive magnetic saturation. Each image details the perpendicular magnetization averaged over the thickness of the film. At remanence (Fig. 1f), the film exhibits stripe domains with a stripe periodicity of 180 nm consistent with Fig. 1c imaged for a full film. The stripe domains are preferentially aligned in the direction of the Fe/Gd wire length. Under a perpendicular field, we observed first coexisting stripe domains and dipole skyrmions (Fig. 1g) and then a closed packed lattice of dipole skyrmions (Fig. 1h) which is similar to the results of Ref. [29]. The dipole skyrmion features are 90 nm in diameter at $H_z = +2000$ Oe. Limitations in the applied magnetic field did not allow the observation of the disordered dipole skyrmion phase and saturation, as previously reported in Ref. [29]. The saturation field for this Fe/Gd wire exists above $H_z > 2800$ Oe, which is above the maximum applied magnetic field of the microscope.

**Current-induced dynamics of stripe domains and dipole skyrmions.**

Figure 2a shows the domain morphology of coexisting stripe domains and dipole skyrmions that form in Pt(5nm)/[Fe(0.34nm)/Gd(0.4nm)]x100/Ta(3nm) film under a perpendicular field of $H_z = +1250$ Oe. We applied a series of 10-*ms*-long current pulses with density amplitude of $j = 5 \times 10^8$ A/m$^2$, along the length of the Fe/Gd wire (see Methods, for further details). We generally observe that the first current pulse alters the domain structures and then subsequent current pulses induce directional motion of the domains. Figures 2b-d details a sequence of snapshots that show the evolution of the domain states when these are subjected to a sequence of single current pulses.

Depending on the current pulse iteration, denoted $p_1$, $p_2$, and $p_3$ in Figs. 2a-d, we find there is a mixture of pinned and moving dipole skyrmions. By tracking the current-induced motion of dipole skyrmions, we observe these textures move in the direction of the current flow (Supp. Movie 1). Example dipole skyrmions that are continuously displaced under the application of the current pulses have been outlined in Figs. 2a-d. When the electrical current is reversed, we observe the dipole skyrmions move opposite to the original direction and again along the current flow direction (Figs. 2e-h). Since the motion is along the current flow direction we attribute the motion arises by SOT from the Pt seed layer acting on the chiral spin structure at the bottom interface and the direction of propagation is consistent with the results of Ref. 16. The top Ta layer is thinner and partially oxidizes and therefore we expect it to contribute far less to the domain motion.

To further probe the driving mechanism that results in current-motion of dipole skyrmions, we have also studied the current dynamics for a Ta(5nm)/[Fe(0.34nm)/Gd(0.4nm)]x100/Pt(3nm) film (Fig. 3) where the materials of the seed and capping layers are inverted. Since heavy transition metals Pt and Ta have opposite sign of the spin Hall angle [33, 34], we expect the motion of dipole skyrmions with respect to the current flow to be reversed if SOT is the driving mechanism. Figure 3a shows the domain morphology, primarily consists of stripe domains and a few dipole skyrmions when the film is exposed to a magnetic field of $H_z$ = -875 Oe (Fig. 3a). After the first current pulse, most of the stripe domains collapse into a collection of dipole skyrmions that occupy the region previously filled by the original stripe domain (Fig. 3b). In the previous magnetic film (Fig. 2) there was limited evidence of stripe domains collapsing into dipole skyrmions. As discussed in Ref. [29], we find that dipole skyrmions can form by two mechanisms: (i) the extremities of a single stripe domain can collapse to form a dipole skyrmion, or (ii) a single stripe domain can pinch into a line of dipole skyrmions where the stripe domain previously existed. Both these mechanisms were previously observed in Fe/Gd films [29] and appear to be favored under different ratios of magnetic energies suggesting these Fe/Gd wires possess small differences in magnetic properties for different seed layers. Figures 3b-e detail subsequent snapshots of the domain morphology after iterative single electrical pulse excitations. Once the film primarily consists of dipole skyrmions, the current-driven stripe to dipole skyrmion transformation is no longer predominant. After the first pulse there is a mixture of pinned and moving dipole skyrmions; however, the displacement of dipole skyrmions is now in the direction of the electron flow (Figs. 3b-e). Examples of moving dipole skyrmions have been outlined in Figs. 3a-e. When the magnetic field is reversed, $H_z = +875$ Oe, we observe comparable current-induced dynamics of the magnetic textures as outlined for $H_z = -875$ Oe, and dipole skyrmions again displace along the electron flow (Fig. 3f-j). As discussed below, this is expected as the chirality of the surface closure domains remains the same after inverting the field. Further imaging of stripe domain response to current pulses including current induced stripe-to-skyrmion generation are shown in Supplementary Section 1 and 2.

**Current-induced motion in dipole skyrmion lattice.**

We also investigated the current-induced dynamics that results from a closed packed lattice of dipole skyrmions using current pulse excitations. Figure 4 shows the domain morphology of Ta(5nm)/[Fe(0.34nm)/Gd(0.4nm)]x100 /Pt(3nm) that forms under a perpendicular magnetic field of $H_z = -2250$ Oe (Fig. 4a). The dipole skyrmion features are 85 nm in diameter and arrange in a disordered closed-pack array. When a current pulse is injected, complex motion is observed throughout the field-of-view (Supp. Movie 3). It is difficult to track individual dipole skyrmions by comparing before/after static images because these textures are arranged in a closed packed lattice making the identification of individual skyrmions difficult. To illustrate dipole skyrmion motion, we have processed raw images, such as the one shown in Fig. 4a, to remove the background and construct binary images. Figure 4b shows the post-processed image that was obtained from Fig. 4a where the white regions are the skyrmions. Changes in the domain morphology with current pulses are outlined as contrast changes that result from the summation of binary images obtained before/after a current pulse is applied. Figures 4c-f show the domain motion that result from four iterative single current pulses where the skyrmions that are pinned appear as white domains in the summation and skyrmions that move will yield an intermediate contrast. We observe there are regions of pinned dipole skyrmions coexisting with moving dipole skyrmions. The displacements of dipole skyrmions are in the form of river-like channels that are oriented at various angles relative to the electron flow.

**Discussion.**

Since the closure domains of a dipole skyrmions possess an opposite Néel-like chiral structure on the top and bottom surface (Fig. 1a, b) and heavy transition metal interfaces with different spin Hall angle, we can expect the coupling to spin currents will be asymmetric at the top and bottom surface. The latter results in a dipole skyrmion experiencing an opposite SOT at each surface. Since the SOT depends on the thickness [36, 37] and the spin-Hall angle [33, 34] of the heavy transition metal, we expect the force amplitude is different at both surfaces. Since the observed directions of the skyrmion motion can be explained by the combination of chirality of the Néel caps and the SOT at the bottom surface, we conclude that the SOT arising from the thicker seed layer plays a dominant role in the motion of dipole skyrmions. Results in Figs. 2 and 3 show there is a dependence in the direction of dipole skyrmion motion relative to inverting the seed/capping layers. A more efficient displacement of dipole skyrmions will likely result when using the same heavy transition metal at both surfaces. Evidence of SOT driving the motion of dipole skyrmions is also supported by the observation of stripe-to-skyrmion generation [38].

Our results on the current-induced motion in the skyrmion lattice phase are similar to a recent numerical study of skyrmion lattices in the presence of disorder [35]. Using a particle-based model of a modified Thiele equation, these authors predicted the skyrmion Hall effect depends on the drive force acting on the skyrmions as well as the pinning force that arises from defects. The study anticipates that: (i) under weak drive forces, the skyrmion Hall angle is not constant and skyrmion motion is oriented in scattered directions; (ii) Increasing the drive force acting on the skyrmions, median drive force, results in the river-like motion that is oriented at an angle relative to the direction of the drive force. (iii) Only after a sufficiently high drive force is applied, do all the skyrmions domains move uniformly along the skyrmion Hall angle. In our Fe/Gd films, we appear, within the context of this model, to be inducing dipole skyrmion motion in the creep regime under a median drive force. We find that reducing the current pulse amplitude to $j = 3 \times 10^8$ A/m results in motion of clustered dipole skyrmions in random regions within the field-of-view, as numerically predicted by Ref. [35] under the application of a weak drive force. In the Supplementary Section 3, we detail current-induced dynamics of a closed packed lattice of dipole skyrmions under the application of a single current pulse with different density amplitude which reveal the transition from the pinned to creep regime. Because of instrumentation limitations, we were unable to inject higher amplitude current pulses (*e.g.* high drive force regime) to observe dipole skyrmion motion in the flow regime. We expect that skyrmion Hall effect [39-42] would be manifested by dipole skyrmions in the flow regime, such that dipole skyrmions with clockwise (Fig. 1a) and counter-clockwise (Fig. 1b) Bloch helicity component would displace tangentially towards opposite direction. In a patterned wire, the latter would result in a net-zero topological and skyrmion Hall effect because the dipole skyrmion accumulation would be equally populated.

Overall, we demonstrated that sub-100nm dipole skyrmions can be manipulated with spin currents via spin-orbit torque. We have experimentally verified the presence of closure domains in our Fe/Gd films which supports the numerically predicted spin structure of dipole skyrmions that form in dipole magnets with $Q$-factor $< 1$. The observation of complex and non-trivial current dynamics resulting from large ensemble of magnetic spin textures illustrates the importance of understanding pinning and repulsive forces. Overall, the ability to manipulate dipole skyrmions with spin currents encourages potential implementation into spintronic based devices and sensors.

**Acknowledgements**

Work a Space and Naval Warfare Systems Center Pacific including nanofabrication and participation at synchrotron experiments was supported by Naval Innovation Science and Engineering Program. Work at University of California San Diego including sample growth and materials characterization and participation at synchrotron experiments was supported by U.S. Department of Energy (DOE), Office of Basic Energy Sciences (Award No. DE-SC0003678). Work at the Advanced Light Source was supported by the U.S. Department of Energy (DE-AC02-05CH11231). Work at the National Institute of Standards and Technology's Center for Nanoscale Science and Technology was performed under project number R13.0004.04. I.G. acknowledges support from the National Research Council's Research Associateship Program. Mi-Young Im acknowledges support by Leading Foreign Research Institute Recruitment Program through the NRF of Korea funded by the MEST (2012K1A4A3053565) and by the DGIST R&D program of the Ministry of Science, ICT and future Planning (18-BT-02).

**Methods**

**Sample Fabrication.** The amorphous Fe/Gd films were fabricated by means of magnetron sputtering at room temperature in a 3-mTorr Ar pressure. The sputter chamber base pressure was $< 3 \times 10^{-8}$ Torr. The Fe/Gd films were grown by alternatively depositing Fe and Gd layers and then repeating the process until a multilayer with the desired number of N repetitions was achieved. The Fe/Gd specimens are deposited onto photolithographic defined wires on 100nm thick SiN membranes with 5x5mm frame and 1mm x 1mm window. Heavy transition metal Pt and Ta seed and capping layers were used to protect the specimen from corrosion. The wire was positioned on the center of the window to enable imaging away from the SiN frame edges. During the deposition, we also simultaneously deposited the Fe/Gd specimens onto Si-substrate with a native-oxide layer for magnetometry measurements and continuous SiN membranes to serve as reference samples.

**Imaging Techniques** The magnetic features at remanence in amorphous Fe/Gd continuous films were imaged using scanning electron microscopy with polarization analysis at the National Institute of Standards and Technology (NIST) in Gaithersburg, MD. The field-dependent magnetic features in amorphous Fe/Gd patterned wires were imaged using full-field soft x-ray transmission microscopy at Lawrence Berkeley National Lab, Advanced Light Source, Beamline 6.1.2 along the Fe $L_3$ (708eV) absorption edge. In order to clearly identify current-induced motion of magnetic textures, an area of the wire with an artifact within the field of view was chosen to facilitate alignment between image snapshots. Each snapshot is the result of a 2 sec exposure and 5 averages.

**Current-induced motion of magnetic domains.** The domain morphology at any given field was recorded prior to the application of an electrical current pulse. For the measurements presented, we used 10-*ms* long current pulse with amplitude of $j = 5 \times 10^8$ A/m². After the current excitation pulse, the domain morphology was again recorded. This process was repeated until collecting several frames was obtained. Afterwards, the magnetic specimen field history was cycled (e.g. the magnetic field was increased to magnetic saturation and then the field was reduced to the opposite magnetic saturation field, prior to returning to the desired magnetic field).

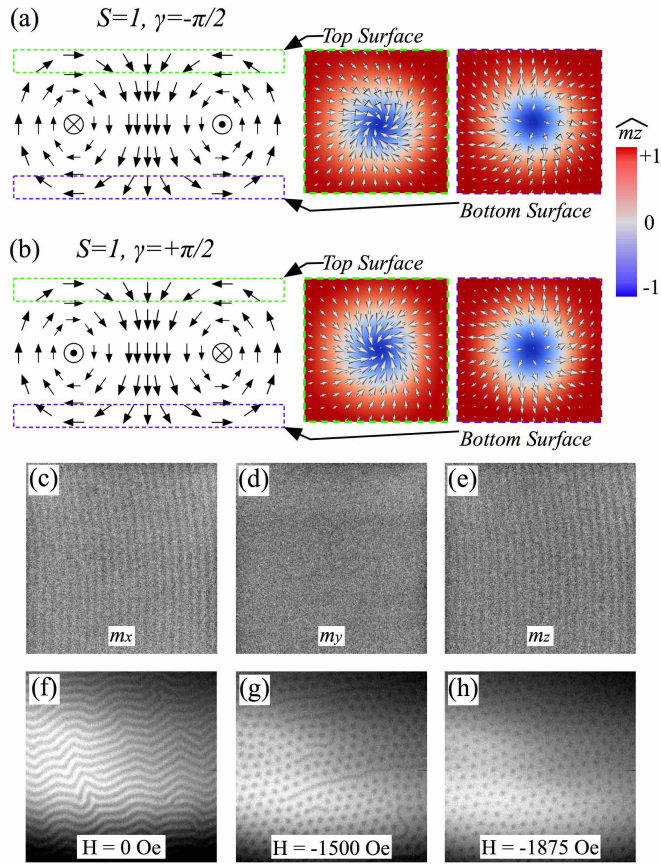

**Figure 1.** Magnetic spin configuration of a dipole skyrmion with (a) clockwise and (b) counter-clockwise helicity. (c-h) Shows the domain morphology of Pt(5nm)/[Fe(0.34nm)/Gd(0.4nm)]x100 /Ta(3nm) at room temperature. (c-e) Detail the surface magnetization ($m_x$, $m_y$, $m_z$) of the stripe domain phase at remanence and (f-h) show average-thickness perpendicular magnetization ($m_z$) field-dependent domain morphology.

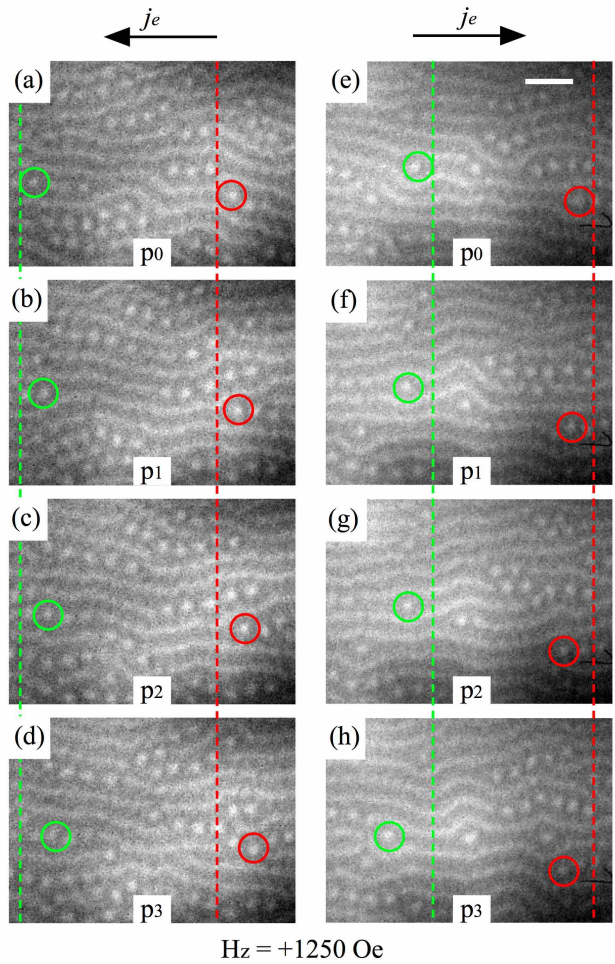

**Figure 2.** Current induced motion observed in coexisting stripe and dipole skyrmion magnetic phase in Pt(5nm)/[Fe(0.34nm)/Gd(0.4nm)]x100/Ta(3nm) at room temperature under a perpendicular magnetic field of $H_z = +1250$ Oe. (a-d) Domain motion that results from applying a positive polarity current, where (a) shows the domain state prior to the current pulse and (b-d) domain states after three iterative current pulses. (e-f) Detail current-driven domain motion for a negative polarity current. Dipole skyrmions are outlined to serve as a guide to the eye to track their motion and the electron flow direction is labeled for both cases. The scale bar in (a) corresponds to 1μm.

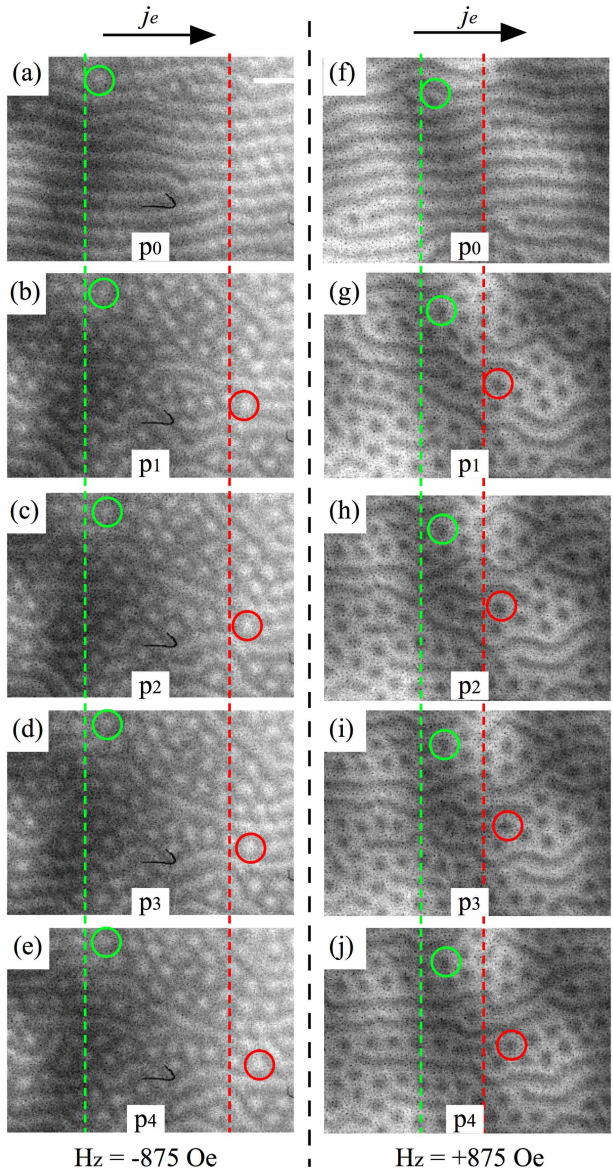

**Figure 3.** Current induced motion observed in coexisting stripe and dipole skyrmion magnetic phase in Ta(5nm)/[Fe(0.34nm)/Gd(0.4nm)]x100/Pt(3nm) at room temperature when exposed to a negative polarity current pulse. (a-e) Domain motion of magnetic states that exist under a perpendicular magnetic field of $H_z = -875$ Oe, where (a) shows the domain state prior to the current pulse and (b-e) domain states after three iterative current pulses. (f-j) Domain motion of magnetic states under $H_z = +875$ Oe, where (f) shows the domain state prior to the current pulse and (g-j) domain states after three iterative current pulses. Dipole skyrmions are outlined to serve as a guide to the eye to track their motion. The scale bar in (a) corresponds to 1μm.

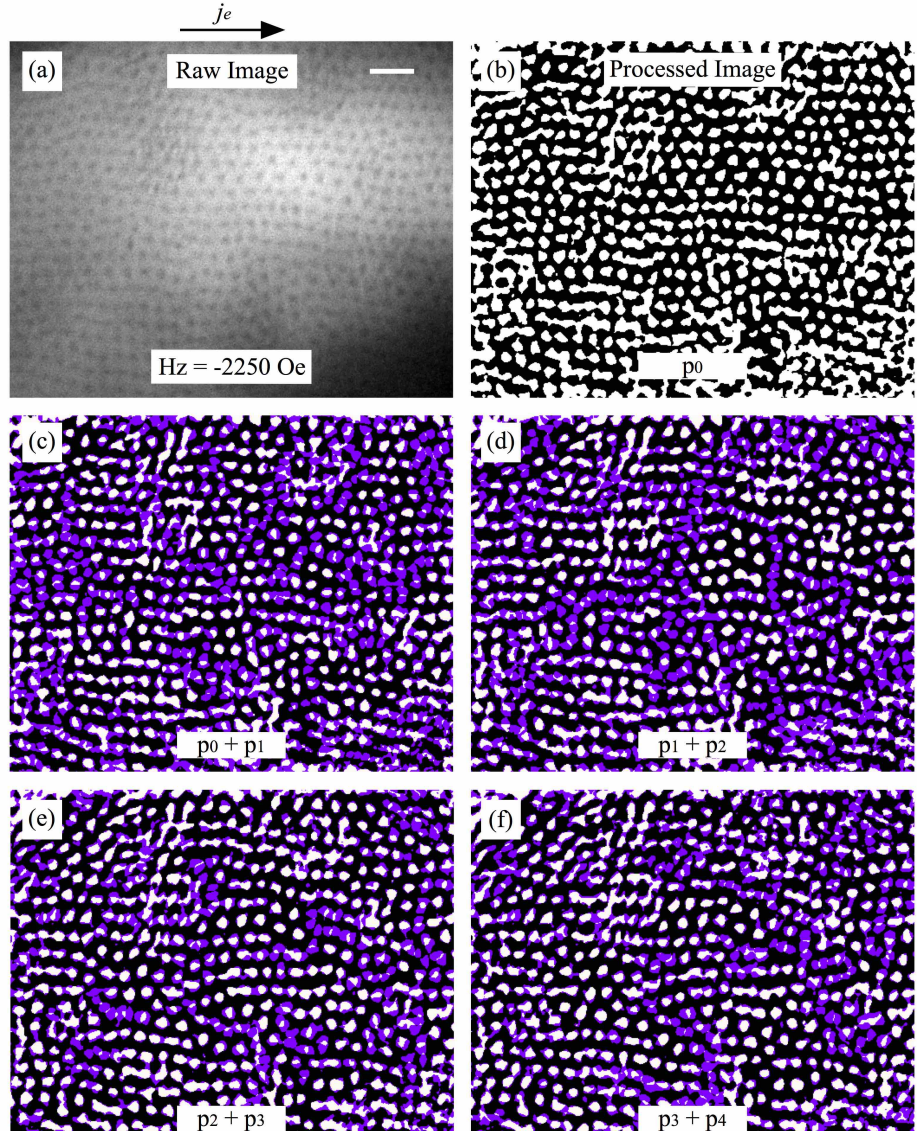

**Figure 4.** Current induced motion of dipole skyrmions arranged in a closed packed lattice in Ta(5nm)/[Fe(0.34nm)/Gd(0.4nm)]x100/Pt(3nm) at room temperature. (a) Domain morphology is obtained under the application of a negative perpendicular magnetic field, $H_z = -2250$ Oe. (b) Post-processed binary image details the domains morphology shown in (a) with background subtracted. (c-d) The dynamics of closed packed dipole skyrmions that result from injecting four consecutive negative polarity current pulses. Each frame details the sum of the images detailing the domain morphology before and after the injecting of the current pulse. The purple colored regions detail regions where changes have occurred in the domain morphology after the current pulse was applied. The scale bar in (a) corresponds to 1μm.

# Supplementary Information

## Spin-orbit torque induced dipole skyrmion motion at room temperature

Sergio A. Montoya[1], Robert Tolley[2,3], Ian Gilbert[4], Soon-Gun Je[5,6], Mi-Young Im[5,6], Eric E. Fullerton[2,3]

[1]Space and Naval Warfare Systems Center Pacific, San Diego, CA 92152, USA
[2]Center for Memory and Recording Research, University of California, San Diego, La Jolla, CA 92093, USA
[3]Department of Electrical and Computer Engineering, University of California, La Jolla, CA 92093, USA
[4]Center for Nanoscale Science and Technology, National Institute of Standards and Technology, Gaithersburg, Maryland 20899, USA
[5]Center for X-ray Optics, Lawrence Berkeley National Laboratory, Berkeley, California 94720, USA
[6]Department of Emerging Materials Science, Daegu Gyeongbuk Institute of Science and Technology, Daegu, Korea

Dated: May 30, 2018

## 1. Current-induced dynamics of stripe domains at remanence.

At remanence, the Fe/Gd multilayer films exhibit perpendicular magnetic stripe domains with a domain period of 180 nm (Supp. Fig. 1a). When the stripe domain phase is exposed to a single 10-*ms* long current pulse with amplitude $j = 5 \times 10^8$ A/m$^2$, we observe the domain morphology becomes altered. To map the domain evolution with current, we have post-processed raw images, as the one detailed in Supp. Fig. 1a, to remove the background and construct binary images (Supp. Fig. 1b). Changes in the domain morphology are outlined as contrast changes that result from the summation of binary images obtained before/after a current pulse is injected. Supplementary Figure 1c-f show the domain reconfiguration that result from injecting four iterative single current pulses, denoted $p_1$, $p_2$, $p_3$ and $p_4$. Notable dynamics are observed in regions where magnetic stripe domain extremities are present including remnant dipole skyrmions and stripe domain dislocations. Since dipole skyrmions are clustered and surrounded by neighboring stripe domains, in this magnetic phase, we do not expect to observe current-induced motion of these textures under these current density amplitudes.

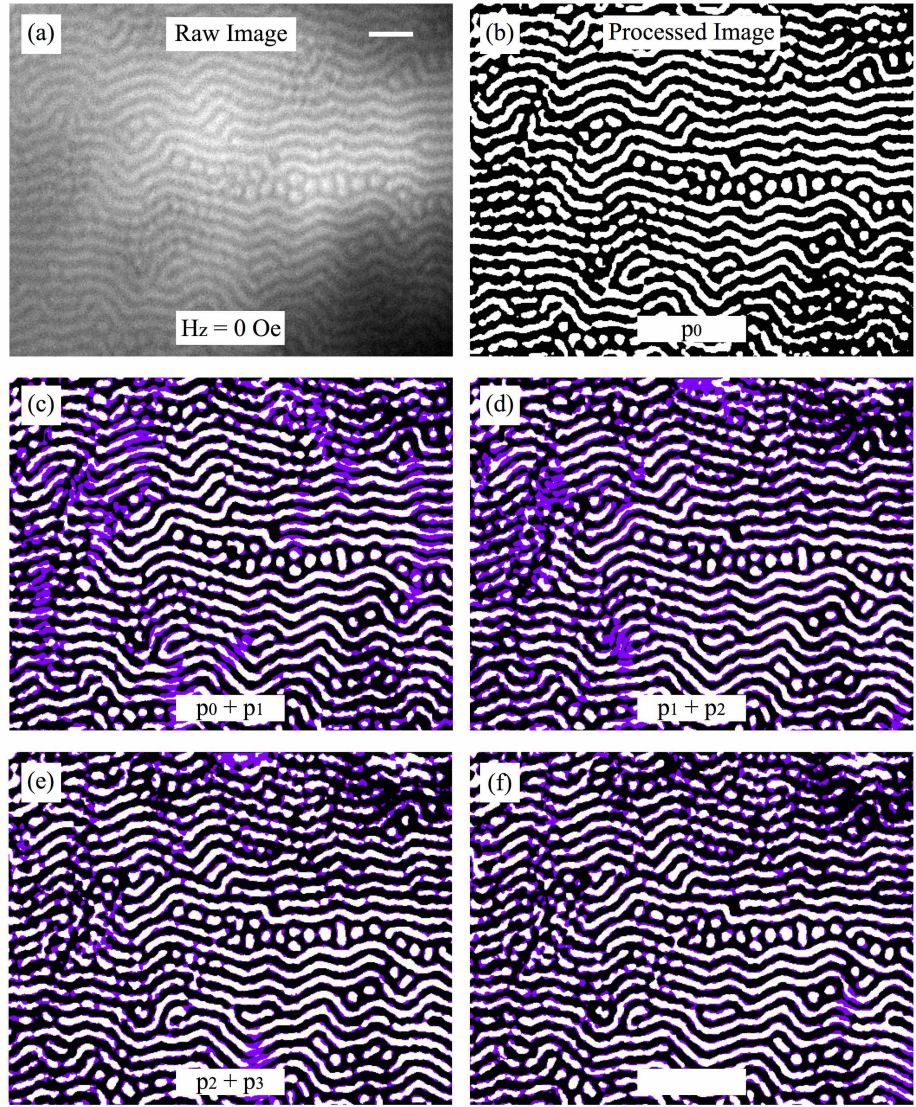

**Supplementary Figure 1**. Current induced dynamics of magnetic stripe domains that form at remanence in Ta(5nm)/[Fe(0.34nm)/Gd(0.4nm)]x100/Pt(3nm) at room temperature. (a) Domain morphology is obtained by means of soft X-ray microscopy measured at the Fe $L_3$ (708eV) absorption edge. (b) Post-processed binary image details the domains morphology shown in (a) with background subtracted. (c-d) The magnetic stripe dynamics that result from injecting a single 10-ms long current pulse with $j = 5\times10^8$ A/m$^2$ are detailed for four current pulses. Each frame details the sum of images detailing the domain morphology before and after injecting a current pulse. The purple colored regions detail areas where changes have occurred in the domain morphology after the current pulse was applied. The scale bar in (a) corresponds to 1μm.

## 2. Current-induced stripe to dipole skyrmion generation.

The field-dependence of current-induced dynamics and stripe to skyrmion generation was investigated in Ta(5nm)/[Fe(0.34nm)/Gd(0.4nm)]x100/Pt(3nm) at room temperature as the sample undergoes a field-induced transition from a stripe domain phase to a coexisting stripe and dipole skyrmion phase. Supplementary Figures 2 and 3 shows the domain morphology before and after it is exposed to a single current pulse of current density amplitude $j = 5 \times 10^8$ A/m$^2$. After injecting the current pulse, the domain morphology is reset by cycling the magnetic field up to positive saturation, then negative saturation and finally applying a higher $H_z$ field than before). In this sample, a stripe domain phase exists from remanence to $H_z = 850$ Oe, and the phase of coexisting stripe and dipole skyrmions spans from $H_z = 850$ Oe to $H_z = 1100$ Oe. We find the current driven stripe to skyrmion generation occurs as we approach coexisting stripe and dipole skyrmion phase. For magnetic fields below $H_z = 625$ Oe, the current pulse excitation primarily rearranges the stripe domains as observed in Supp. Figs. 2a-f. For $H_z \geq 625$ Oe, the current pulse begins to transform stripe domains to dipole skyrmions and the generation of dipole skyrmion increases with applied perpendicular field (Supp. Figs. 2g-i and 3). At magnetic fields where the current-induced dynamics and stripe-to-skyrmion generation is less evident, we illustrate changes between the initial and final domain states through the sum of postprocessed binary of images of the latter, such that contrast changes detail the areas where the domain morphology becomes altered (Supp. Figs. 2c, f, i).

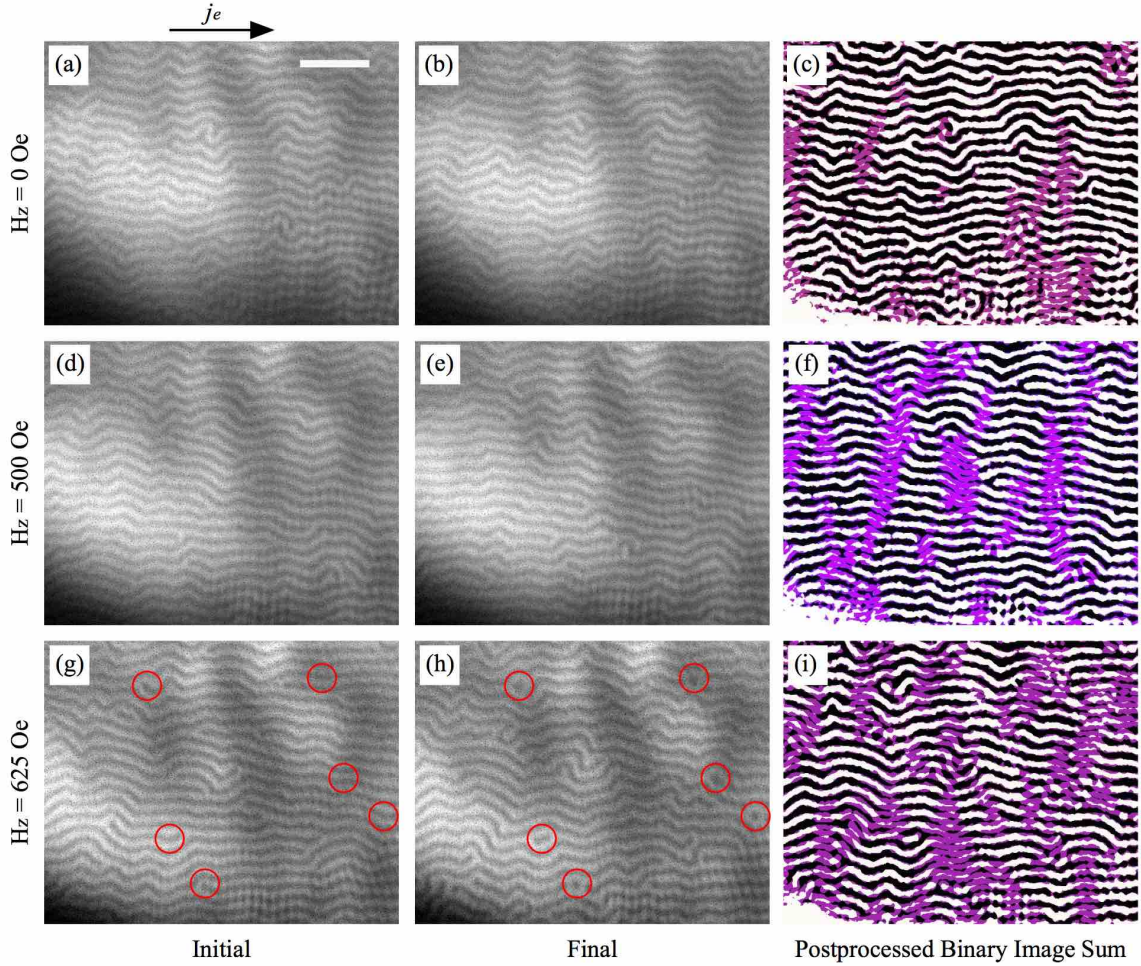

**Supplementary Figure 2.** Current-induced dynamics and stripe domain to dipole skyrmion generation in Ta(5nm)/[Fe(0.34nm)/Gd(0.4nm)]x100/Pt(3nm) at room temperature from remanence to $H_z = +625$ Oe. Images detailing the field-depentn domain morphology are obtained by means of soft X-ray microscopy measured at the Fe $L_3$ (708eV) absorption edge. Contrast ripples in the images are artifacts from the X-ray optics. The first column (a, d, g) illustrates the domain morphology before a current pulse is injected at three fields $H_z = 0$, +500 and +625 Oe, whereas the second column (b, e, h) shows the domain states after the current pulse is injected. The third column (c, f, i) details the postprocessed binary image sum of initial and final states. Purple-hue contrast areas represent regions where the current-pulse has modified the domain morphology. At $H_z = +625$ Oe there is evidence of stripe-to-skyrmion generation in scattered regions of the field of view (g, h). Markers have been placed in (g, h) to serve as guide to the eye. The scale bar in (a) corresponds to 1µm.

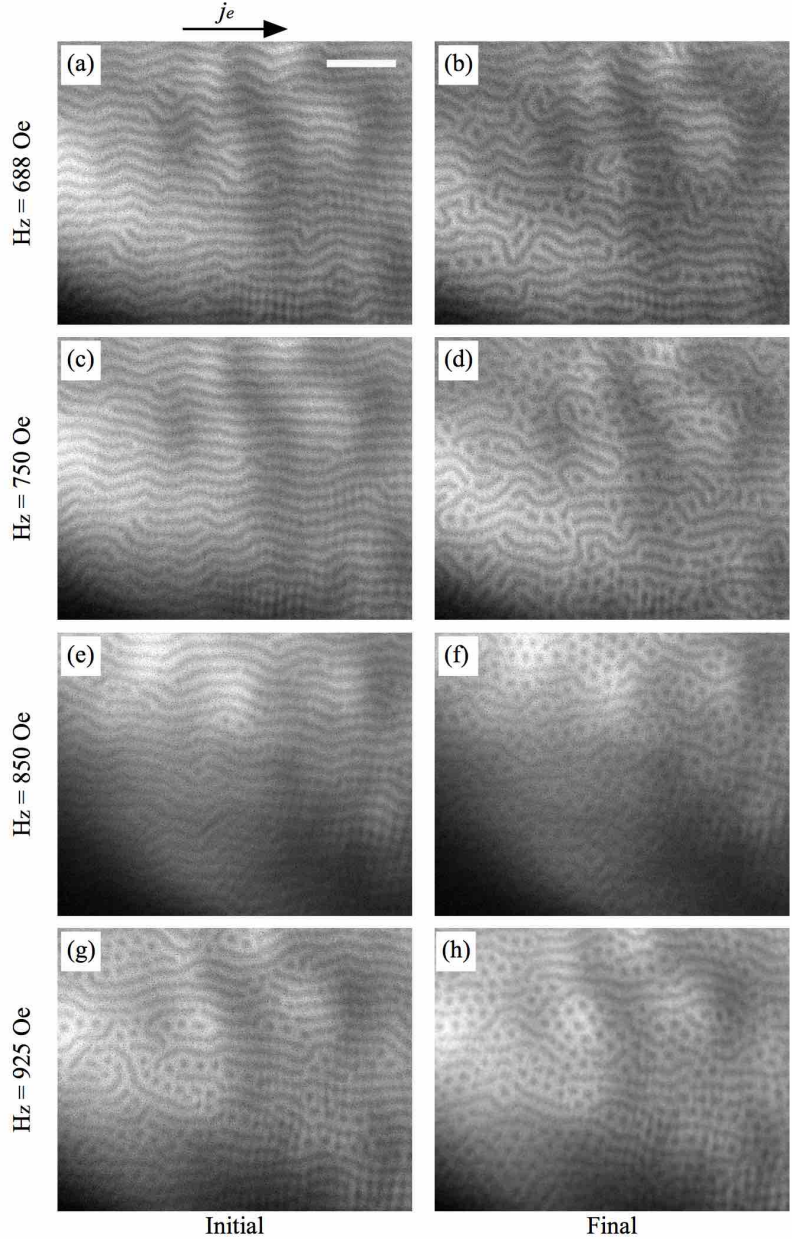

**Supplementary Figure 3.** Current-induced stripe domain to dipole skyrmion generation in Ta(5nm)/[Fe(0.34nm)/Gd(0.4nm)]x100/Pt(3nm) at room temperature from $H_z = +688$ Oe to $H_z = +925$ Oe. Images detailing the field-dependent domain morphology are obtained by means of soft X-ray microscopy measured at the Fe $L_3$ (708eV) absorption edge. Contrast ripples in the images are artifacts from the X-ray optics. The first column (a, c, e, g) shows the initial domain morphology state as a function of increasing perpendicular field, and the second column (b, d, f, h) shows the domain states after the injection of the single current pulse. The scale bar in (a) corresponds to 1µm.

3. **Dipole skyrmion motion dependence on the current density.**

We explored the current-induced motion of dipole skyrmions arranged in a closed packed lattice under different current density amplitudes. To highlight the motion of the skyrmions the domain images are converted into binary images and motion is outlined as contrast changes that result from the summation of binary images obtained before/after current pulses are applied. After a series of current pulses are injected, at a current density amplitude, the domain morphology is reset by cycling the perpendicular field up to positive saturation, then reducing the field to negative saturation, and finally increasing the field up to $H_z = +2000$ Oe under which a closed-packed lattice of dipole skyrmions forms. In our experiments, we observe the transition from the pinned to creep regime as we probe dipole skyrmions with increasing current density amplitudes from $j = 0.5 \times 10^8$ A/m$^2$ to $j = 5 \times 10^8$ A/m$^2$ (Supp. Fig. 4). Current amplitudes below $j < 3 \times 10^8$ A/m$^2$ exerts a weak SOT that is unable to displace the textures from their pinning site. Supplementary Figures 4a-c show the domain states after the injection of three iterative current pulses with amplitude $j = 0.5 \times 10^8$ A/m$^2$, where there is evidence of minute changes in contrast around many of the individual textures. The latter results from illumination artifacts when recording domain morphology snapshots. Increasing the current density amplitude, $j > 3 \times 10^8$ A/m$^2$, we observe the number of dipole skyrmions exhibiting motion increases from a few clustered textures in scattered regions across the field-of-view to many textures that form river-like channels (Supp. Fig. 4d-l).

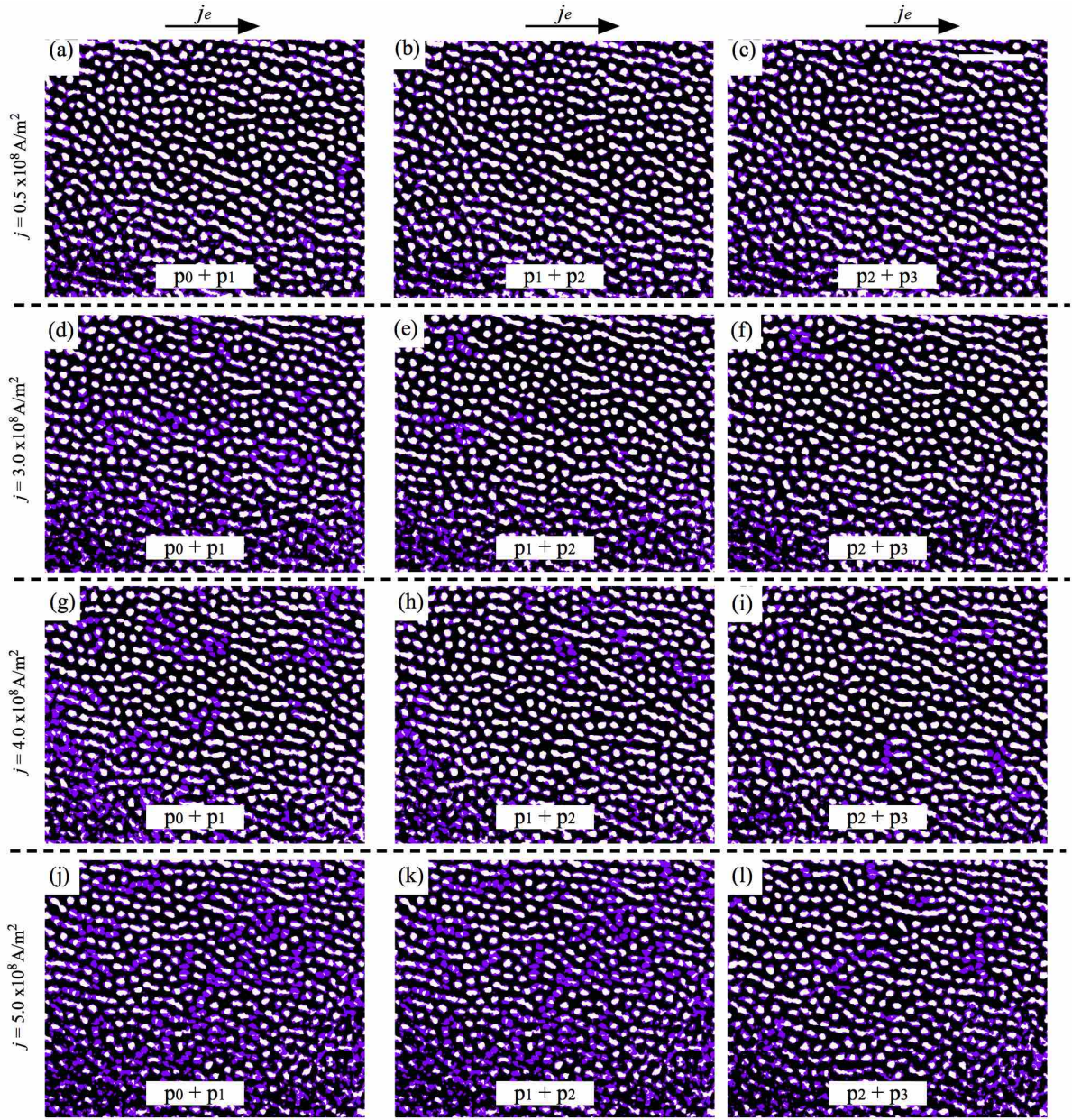

**Supplementary Figure 4**. Current-induced motion of dipole skyrmions arranged in a closed packed lattice in Ta(5nm)/[Fe(0.34nm)/Gd(0.4nm)]x100/Pt(3nm) at room temperature. Each frame shows the sum of the binary images detailing the domain morphology before/after injecting current pulse excitations with different current density amplitude: (a-c) $j = 0.5 \times 10^8$ A/m², (d-f) $j = 3 \times 10^8$ A/m², (g-c) $j = 4 \times 10^8$ A/m² and (j-l) $j = 5 \times 10^8$ A/m². At each current density, the dynamics are captured for three consecutive current pulses. The purple colored regions show areas where changes have occurred after the injection of a current pulse. The scale bar in (c) corresponds to 1µm.